\documentclass[aps,prb,onecolumn,floatfix]{revtex4}

\usepackage{epsf}
\usepackage{subfigure}
\usepackage{amsmath}
\usepackage{amssymb}
\usepackage{bm}
\usepackage{graphicx}
\usepackage{epstopdf}
\newcommand{\be}{\begin{equation}}
\newcommand{\ee}{\end{equation}}
\newcommand{\bea}{\begin{eqnarray}}
\newcommand{\eea}{\end{eqnarray}}

\newcommand{\parent}[1]{\left( #1 \right)}

\begin{document}

\title{\bf Equivalence classes for large deviations}

\author{David Andrieux}

\begin{abstract}
We show the existence of equivalence classes for large deviations. 
Stochastic dynamics within an equivalence class share the same large deviation properties. 
\end{abstract}

\maketitle

\vskip 0,25 cm

\section*{KEY TAKEAWAYS}

\begin{itemize}
%
%
%
%
%

\item An equivalence relation for stochastic dynamics is defined. Stochastic dynamics that belong to the same equivalence class have the same large deviations.

%
\item Stochastic dynamics can be factored into an equilibrium and a nonequilibrium part. The equilibrium part is common to all elements within an equivalence class.
\item	Large deviations can be expressed in terms of a scalar field defined on the equivalence classes. The symmetry of this field gives rise to the symmetries of the fluctuations theorems.
\item The large deviations of an equilibrium dynamics determine the large deviations of all the nonequilibrium dynamics within its equivalence class.

\end{itemize}


\vskip 0,75 cm

\section{Framework}

\subsection{Markov chains}

We consider a Markov chain characterized by a transition matrix $P = \parent{P_{ij}} \in \mathbb{R}^{N\times N}$ on a finite state space.
The operator $P$ is stochastic, i.e., it is non-negative ($P \geq 0$) and its rows sum to one ($\sum_{j} P_{ij} = 1)$.
We assume that the Markov chain is primitive, i.e., there exists an $n_0$ such that $P^{n_0}$ has all positive entries. 
This guarantees that $P$ has a unique stationary distribution $\pmb{\pmb{\pi}}$ such that $\pmb{\pmb{\pi}} = \pmb{\pmb{\pi}} P$.

For further reference, we note that a non-negative matrix can be transformed into a stochastic matrix. 
Indeed, consider a non-negative matrix $A$ and denote its Perron root by $\chi$ and its right Perron vector by $x$. Then if $D = {\rm diag}(x_1, \ldots, x_N)$ we have that
\begin{eqnarray}
P = \frac{1}{\chi} D^{-1} A D
\label{mapT}
\end{eqnarray}
is a stochastic matrix.

\subsection{Thermodynamic description}

A Markov chain $P$ defines an equilibrium dynamics when
\bea
P_{i_1 i_2} \ldots P_{i_n i_1} = P_{i_1 i_n} \ldots P_{i_2 i_1 } \quad  \text{for any finite sequence} \; (i_1, i_2, \ldots, i_n) \, .
\label{Kolmogorov}
\eea
In this case, no probability flux is present at the stationary state.

When the conditions (\ref{Kolmogorov}) are not satisfied we are in presence of a nonequilibrium dynamics. 
In this case, probability fluxes ciculate through the system and generate thermodynamic forces or affinities. 
The affinities can be measured by the breaking of detailed balance along cyclic paths $c = (i_1, i_2, \ldots, i_n)$ as
\bea
\frac{P_{i_1 i_2} \ldots P_{i_n i_1}}{ P_{i_1 i_n} \ldots P_{i_2 i_1 } } = \exp \parent{A_c} \, .
\label{A}
\eea

However, Schnakenberg demonstrated that only a subset of these affinities are independent.\cite{S76} 
He also showed how to identify a basis of affinities $\pmb{A} = (A_1, \ldots, A_l, \ldots, A_M)$ based on the concept of fundamental chords. 
A summary of his theory is presented in the Appendix.

\section{Thermodynamic decomposition of stochastic dynamics}

Stochastic dynamics can be brought into a "thermodynamic form" through a similarity transform. 
This form reveals the thermodynamic conditions of the dynamics: It is symmetric at equilibrium, while the symmetry is explicitly broken by the affinities out of equilibrium.

The thermodynamic form is expressed in terms of two operators $E$ and $Z$, defined as
\bea
E_{ij} [P] = \sqrt{P_{ij} P_{ji} } 
\eea
and
\bea
Z_{ij} (\pmb{\mu}) \equiv \begin{cases} \exp \parent{\pm \mu_l/2}  & \text{if the transition $i\rightarrow j$ is a fundamental chord $l$ in the positive (negative) orientation,} \\
                1 & \text{otherwise}.
                \end{cases}
\eea
The operator $E$ is symmetric, $E = E^{{\rm T}}$, while $Z(\pmb{\mu}) = Z^{{\rm T}} (- \pmb{\mu})$.\\


{\bf Proposition 1.} {\it A stochastic matrix $P$ with affinities $\pmb{A}$ is similar to 
\bea
E[P] \circ Z (\pmb{A}) \, ,
\label{thermoform}
\eea
where $X \circ Y$ denotes the Hadamard product: $\parent{X\circ Y}_{ij} = X_{ij} Y_{ij}$.}\\

DEMONSTRATION:  
We consider the similarity transform 
\bea
P' = U P U^{-1}
\nonumber
\eea
with $U = {\rm diag} (u_1, \ldots, u_N )$. Therefore $P^{'}_{ij} = P_{ij} (u_i /u_{j})$. 
We choose the elements $u_i$ such that 
\bea
\frac{u_{i}}{u_{j}} = \left( \frac{P_{ji}}{P_{ij}} \right)^{\frac{1}{2}}
\label{uratio}
\eea
for all the transitions $i \rightarrow j$ that do not correspond to a chord. 
We have thus constrained the ratios (\ref{uratio}) along the maximal tree. 
By construction, this provides a consistent set of equations whose solution is determined up to a multiplicative factor.

The elements of $P^{'}$ then take the values
\bea
P^{'}_{ij} = \sqrt{P_{ij} P_{ji}} = E_{ij} [P] \, Z_{ij}
\label{Mnochord}
\nonumber
\eea
if the transition $i \rightarrow j$ is not a chord. 

The elements corresponding to chords are obtained as follows. 
Consider the chord $l \equiv i_l \rightarrow i_1$ (in the positive orientation) and its fundamental cycle $c_l = (i_1,\cdots,i_l)$. 
We have the identity
\bea
\prod_{k=1}^l \frac{u_{i_{k+1}}}{u_{i_{k}}} = \frac{u_{i_1}}{u_{i_{l}}} \prod_{k=1}^{l-1} \frac{u_{i_{k+1}}}{u_{i_{k}}}  = 1 \, ,
\label{idenprod}
\nonumber
\eea
where $i_{l+1}\equiv i_1$.
By construction, a fundamental cycle $c_l$ only contains its associated chord $l$. Hence, using formulas (\ref{A}) and (\ref{uratio}), we have
\bea
\frac{u_{i_l}}{u_{i_{1}}} = \parent{\frac{P_{i_1 i_l}}{P_{i_l i_1}}}^{\frac{1}{2}}  \prod_{k=1}^{l} \parent{\frac{P_{i_{k}i_{k+1}}}{P_{i_{k+1}i_{k}}}}^{\frac{1}{2}} = \parent{\frac{P_{i_1 i_l}}{P_{i_1i_l}}}^{\frac{1}{2}} {\rm e}^{A_l/2} \, .
\nonumber
\eea
Accordingly, the operator element associated with the chord $l$ reads
\bea
P^{'}_{i_{l}i_{1}}  = \sqrt{P_{i_li_1} P_{i_1i_l}} \ {\rm e}^{A_l/2} = E_{i_{l}i_{1}} [P] Z_{i_{l}i_{1}} (\pmb{A})\, .
\nonumber
\eea
Similarly, 
$P^{'}_{i_{1}i_{l}}  = \sqrt{P_{i_li_1} P_{i_1i_l}} \ \exp \parent{- A_l/2} = E_{i_{1}i_{l}} [P] Z_{i_{1}i_{l}} (\pmb{A}) $
for the negative orientation.
$\Box$\\ 

We call (\ref{thermoform}) the {\it thermodynamic form} of $P$. The operator $E$ captures the equilibrium part of the dynamics while $Z$ reveals its thermodynamic conditions.

Note that this form preserves the eigenvalues of $P$ as well as its affinities. Also, the same similarity transform $U$ brings a {\it non-negative} matrix into its thermodynamic form.

\section{Equivalence classes}

The thermodynamic decomposition (\ref{thermoform}) structures the space of stochastic dynamics into equivalence classes.
We define the {\it equivalence relation} 
\bea
P \sim H \quad \text{if there exists a factor} \ \gamma \ \text{such that} \quad E[P] = \gamma  \ E[H] \,  .
\label{eqrel}
\eea
This relation is reflexive, symmetric, and transitive. 
It thus forms a partition of the space of stochastic dynamics: every dynamics belongs to one and only one equivalence class.\\

We can parametrize the elements of an equivalence class. 
For simplicity, we work in the thermodynamic representation (\ref{thermoform}) and use the equivalence \eqref{mapT} between non-negative and stochastic matrices. 
\footnote{A non-negative matrix $A$ and its associated stochastic operator \eqref{mapT} have the same thermodynamic form, up to a multiplicative factor.}\\


{\bf Theorem 1.} {\it The equivalence class $[P]$ is composed of all the dynamics
\bea
E[P] \circ Z(\pmb{\mu}) \, ,
\label{param.[P]}
\eea
where the parameters $\pmb{\mu}$ can take any values in $\mathbb{R}^{M}$.}\\

DEMONSTRATION:
We first show that all elements in $[P]$ can be written in the form (\ref{param.[P]}). This results from Proposition $1$ and the definition of $Z$. 
Conversely, elements of the form \eqref{param.[P]} belong to $[P]$. $\Box$\\

Note that an equivalence class covers all possible thermodynamic conditions. 
Among these, the dynamics $E$ is the unique equilibrium dynamics; it also defines the common structure for all elements in the class.\\

In the following we show that the equivalence relation (\ref{eqrel}) plays a pivotal role in understanding large deviations.

\section{Large deviations and thermodynamic form}

We define a scalar field $\rho$ over the equivalence class $[P]$ as follows: 
\bea
\rho(P, \pmb{\mu}) = \text{spectral radius of} \; \, E[P] \circ Z (\pmb{\mu}) \, .
\eea
It has the symmetry property
\bea
\rho (P, \pmb{\mu}) = \rho(P,-\pmb{\mu})\, .
\label{sym.rho}
\eea 

Large deviations can be expressed in terms of the field $\rho$. 
Indeed, the large deviations of $P$ are obtained as the largest eigenvalue of
\bea
P \circ Q(\pmb{\lambda}) \, ,
\label{LD}
\eea
where $Q (\pmb{\lambda})$ is a non-negative operator that measures the physical quantities of interest. 
Using that $E[Q] = I$, where $I$ is the identity matrix, and that $Z(\pmb{x}) \circ Z(\pmb{y}) = Z(\pmb{x+y})$, the thermodynamic form of (\ref{LD}) reads
\bea
E[P] \circ Z[\pmb{A} + \pmb{q} (\pmb{\lambda})] \, .
\eea
The function $\pmb{q}$ is readily calculated by looking at the transition probabilities of $Q$ along cycles. 
The large deviations are then given by (minus the logarithm of) $\rho [P, \pmb{A} + \pmb{q} (\pmb{\lambda}) ]$.\\

{\bf Example:} {\it Large deviations of the entropy production.} In this case the operator $Q (\lambda)$ takes the form
\bea
Q_{ij}(\lambda) = \parent{\frac{P_{ji}}{P_{ij}}}^{\lambda} \, .
\nonumber
\eea
In the thermodynamic representation $P \circ Q(\lambda)$ reads 
\bea
E [P] \circ Z[(1-2\lambda) \pmb{A}] \, .
\nonumber
\eea
Therefore, the large deviations of the entropy production effectively probe the dynamics in $[P]$ with effective affinities $(1-2\lambda) \pmb{A}$.
They are thus given by (minus the logarithm of) $\rho [P, (1-2\lambda) \pmb{A}]$. 
Note that the fluctuation symmetry $\lambda \rightarrow 1-\lambda$ follows from the symmetry (\ref{sym.rho}).


\section{Dynamics within an equivalence class share their large deviations}


For concreteness, we consider the large deviations of the thermodynamic currents. 
They are measured by the operator $Q (\pmb{\lambda}) = Z (-2\pmb{\lambda})$. 
Therefore, they are obtained from $\rho (P, \pmb{A}-2\pmb{\lambda})$. 
Note that the fluctuation symmetry $\pmb{\lambda} \rightarrow \pmb{A}-\pmb{\lambda}$ follows from the symmetry (\ref{sym.rho}).\\


{\bf Theorem 2.} {\it Dynamics within an equivalence class share their large deviations.} \\

DEMONSTRATION: Note that $\rho (P, \pmb{\mu}) \propto \rho (P', \pmb{\mu})$ if $P \sim P'$. 
Therefore, the large deviations of $P$,  $\rho (P, \pmb{A}-2\pmb{\lambda})$, are related to the large deviations of $P'$, $\rho (P', \pmb{A}'-2\pmb{\lambda}')$, by a multiplicative factor  and the parameter transformation $\lambda' \rightarrow \lambda + (\pmb{A}'-\pmb{A})/2$. 
$\Box$\\

{\bf Corollary 1.} {\it 
The large deviations of the equilibrium dynamics $E[P]$ determine the large deviations of the whole equivalence class $[P]$.
}\\

The equilibrium dynamics $E[P]$ thus shares the same large deviation properties as $P$. 
In this sense, $E[P]$ is the natural equilibrium state of $P$ from a {\it dynamical} viewpoint. \footnote{It does not, however, necessarily coincide with the thermodynamic equilibrium of a parametric model. For example, consider a three-states dynamics $P_a$, with $P_{11} = P_{22} = P_{33} = 0, P_{12} = P_{13} = P_{21} = P_{23} = 1/2, P_{31} = a, P_{32} = 1-a$.
Its thermodynamic equilibrium, from a {\it parametric} viewpoint, occurs when $a=1/2$. However, $E[P_a]$ is not similar to $P_{1/2}$ when $a \neq 1/2$.}

\vskip 1,6 cm

{\bf Disclaimer.} This paper is not intended for journal publication. 

\vskip 0,7 cm

\appendix

\section{Schnakenberg theory}

Schnakenberg's theory decomposes the thermodynamical properties of stochastic dynamics into their independent contributions \cite{S76}.

A graph $G$ is associated with a stochastic dynamics as follows: each state $i$ of the system corresponds to a vertex or node while the edges represent the different transitions
$i\rightleftharpoons j$ allowed between the states.

An orientation is given to each edge of the graph $G$. The directed edges are thus defined by 
$ \equiv i \rightarrow j $.

A graph $G$ usually contains cyclic paths. 
However, not all such paths are independent. 
They can be expressed by a linear combination of a smaller subset of cycles, called the {\it fundamental set}, which plays the role of a basis in the space of cycles. 
To identify all the independent cycles of a graph we introduce a {\it maximal tree} $T(G)$, which is a subgraph of $G$ that satisfies the following properties:
\begin{itemize}
\item  $T$ contains all 
the vertices of $G$;

\item{ $T$ is connected;}

\item $T$ contains no circuit, i.e., no cyclic sequence of edges.
\end{itemize}
In general a given graph $G$ has several maximal trees. 

The edges $l$ of $G$ that do not belong to $T$ are called the {\it chords} of $T$. For a graph with $N$ vertex and $E$ edges, there exists $M=E-N+1$ chords.
If we add to $T$ one of its chords $l$, the resulting subgraph $T+l$
contains exactly one circuit, $c_l$, which is obtained from $T+l$ by removing
all the edges that are not part of the circuit.  
Each chord $l$ thus defines a unique cycle $c_l$ called a {\it fundamental cycle}.
In this paper, we use the convention that the orientation is such that the cycles are oriented as the chords $l$.

We can formulate many important thermodynamic concepts in terms of cycles. 
For instance, the affinity of an arbitrary cycle $c$ can be expressed as a linear combination of the affinities of a fundamental set:
\bea
A(c) = \sum_l \epsilon_l (c) A(c_l) \, ,
\label{Adecom}
\nonumber
\eea
where the sum extends over all the chords, and $\epsilon_l (c) = \pm 1$ if $c$ contains the edge $l$ in the positive ($+$) or negative orientation ($-$), and $0$ otherwise. \footnote{Accordingly, the maximal tree $T$
can be chosen arbitrarily because each cycle $c_l$ can be redefined by
linear combinations of the fundamental cycles.}
The fundamental cycles thus constitute a basis identifying the {\it independent} contributions to the stochastic process.

\newpage


\end{document}